\documentclass[12pt,a4paper]{article}
\usepackage{amsmath}
\usepackage{amssymb}
\usepackage[final]{showkeys}
\usepackage[dvips]{graphicx}

\newcommand{\Cx}{\mathbb{C}}

\newcommand{\Ir}{\mathbb{Z}}
\newcommand{\Nl}{\text{I\hspace{-.4ex}I\hspace{-.7ex}N}}

\def\en{\mathsf{h}}

\def\En{\mathsf{H}}

\def\idty{{\leavevmode{\rm 1\ifmmode\mkern -5.4mu\else
                                            \kern -.3em\fi I}}}

\newcommand{\<}{\langle}
\renewcommand{\>}{\rangle}

\newcommand{\ind}[1]{\null}

\newcommand{\rank}{\mathop\mathrm{Rank}\nolimits}

\renewcommand{\c}[1]{\mathcal{#1}}
\newcommand{\g}[1]{\mathfrak{#1}}

\renewcommand{\r}[1]{\mathrm{#1}}

\def\idty{{\leavevmode{\rm 1\ifmmode\mkern -5.4mu\else %
                        \kern -.3em\fi I}}}
\newcommand{\tr}{\mathop\mathrm{Tr}\nolimits}
\newenvironment{proof}{\setlength{\parindent}{0pt}{\bf Proof:} \par }{\par %
  \hfill $\blacksquare$ \par}
  {\setlength{\parindent}{0pt}{\bf Proof:} \par }%
  {\par \hfil  \par}

\newtheorem{proposition}{Proposition}[section]
\newtheorem{lemma}{Lemma}[section]
\newtheorem{theorem}{Theorem}[section]

\begin{document}
\ \vskip 1cm

\centerline{\LARGE \bf Coherent transport and dynamical entropy} 
\smallskip 
\centerline{\LARGE \bf for Fermionic systems}
\bigskip

\centerline{\large R.~Alicki$^1$, M.~Fannes$^{\,2}$,
B.~Haegeman$^{2,\,}$\footnote[3]{Research Assistant of the Fund for 
Scientific Research - Flanders (Belgium)(F.W.O. - Vlaanderen)} and 
D.~Vanpeteghem$^{2,\,3}$}
\bigskip

\centerline{$^1$\,Institute of Theoretical Physics and Astrophysics}
\centerline{University of Gda\'nsk, PL-80-952 Gda\'nsk, Poland}
\bigskip

\centerline{$^2$\,Instituut voor Theoretische Fysica}
\centerline{K.U. Leuven, B-3001 Leuven, Belgium}
\bigskip\bigskip

\noindent
\textbf{Abstract}
This paper consists in two parts. First we set up a general scheme of local
traps in an homogeneous deterministic quantum system. The current of
particles caught by the trap is linked to the dynamical behaviour of the
trap states. In this way, transport properties in an homogeneous
system are related to spectral properties of a coherent dynamics. Next we
apply the scheme to a system of Fermions in the one-particle approximation.
We obtain in particular lower bounds for the dynamical entropy in terms of
the current induced by the trap.  
\bigskip


\section{Introduction}

In this paper, we are interested in time scaling properties of propagation
by coherent quantum dynamics. Non-trivial behaviour of a  single-particle
Hamiltonian can lead to a description of anomalous diffusion of electrons
in solids~\cite{bel}. This behaviour becomes apparent through the scaling of the
spreading of one-particle wave functions with respect to time. We shall
here adopt another approach: we introduce a localised trap in an
infinite system and study the time behaviour of the current of particles
falling in the trap. Applied to systems of Fermions in the non-interacting
approximation, we obtain a lower bound on the dynamical entropy in terms of
this current. 

The trap states will be described by a collection of wave functions and we
relate the current to the dynamics of these states. In particular, we show
that an absolutely continuous spectrum produces a non-zero asymptotic
current. A singular spectrum will lead to asymptotically vanishing currents
possibly characterised by a dynamical exponent.  

A number of related topics and models have been considered in the
literature, mostly for the case of a continuous time evolution. The
occurrence of singular continuous spectra as a source of anomalous diffusion 
is caused either by randomness in the Hamiltonian or by
aperiodicity. We shall however not be concerned by producing such models
but rather link scaling properties of dynamical entropy to assumed spectral
properties of a discrete dynamics.  

Coherent transport in quantum systems is being studied by using reservoirs
as drivers. A number of delicate questions arise in this context with
respect to the thermodynamic limit. Anomalous transport due to spatial
randomness in the dynamics seems to occur~\cite{woj}.

Strongly chaotic classical or quantum dynamical systems generate entropy at
a non-zero asymptotic rate: the dynamical entropy. In the classical case, 
the sum of the positive Lyapunov exponents is a bound for the entropy
(Ruelle's inequality) and equality is reached for sufficiently smooth
systems (Pesin's theorem). For quantum dynamical systems several entropies
have been introduced such as the CNT~construction based on a coupling with
a classical system and the ALF~construction that relies on POVM's
(operational partitions of unity). In order to obtain a non-zero entropy an
absolutely continuous dynamical  spectrum is needed, at least for Fermion
systems in the one-particle approximation~\cite{sto}. In open classical systems, the
escape rate formalism links the escape exponent from an unstable repeller
to diffusive transport. The  escape rate is given by the missing exponents
in the entropy for motion on the repeller~\cite{gas}. 

There are however many mixing dynamical systems with less pronounced
randomising properties which are not given in terms of exponents or rates.
Such dynamics may lead to a sublinear scaling for the total dynamical
entropy~\cite{ben}. 

The aim of this work is to establish a lower bound for the entropy in terms
of dynamical exponents of a localised trap in an infinite system both in
the regular and the anomalous case.

As a motivation we provide in Section~\ref{section1} a few simple examples 
of the use of a trap in classical dynamics. Obviously there is a  different
physical mechanism at work with possibly similar macroscopic effects but
our aim is to show that the time behaviour of the current at the edge of a
localised trap encodes relevant information about the transport properties
of the dynamics. Because of the locality of the trap, it is possible to
deal immediately with an infinite system and to avoid using a delicate
large volume limit of boundary conditions. Section~\ref{section2} deals
with localised traps in  unitary Hilbert space dynamics and we study in
Section~\ref{section3} the dynamical entropy for non-interacting Fermions.
We obtain in particular a lower bound for the entropy in terms of the
current of particles falling in a trap.     

\section{Traps in a classical context}
\label{section1}

Consider first the simplest classical counterpart of a trap absorbing free
electrons moving with velocities below that corresponding to a given Fermi
level. This is a system of classical particles homogeneously distributed in
space and with uniform velocity distribution below a maximal one. The state
of such a system is a uniform mixture of spatially homogeneous states with
fixed velocity. The number of particles per time unit caught in a localised
trap is constant in time and essentially determined by the average
cross-section of the trap. 

Next we consider the model of the diffusion equation in one dimension with
a trap at the origin. We have to find the solution of the equation
\begin{equation*}
 \frac{\partial n}{\partial t} = D\, \frac{\partial^2n}{\partial x^2}
\end{equation*}  
for $x > 0$ and $t > 0$ with boundary condition $n(0,t)=0$ and initial
condition $n(x,0)=1$. In this equation $D$ is the diffusion constant and
$n(x,t)$ represents the particle density at time $t$ and place $x$ and the
particle current is given by Fick's law
\begin{equation*}
 j(x,t) = - D \frac{\partial\ }{\partial x} n(x,t).
\end{equation*}   
The solution of the diffusion equation reads
\begin{equation*}
 n(x,t) = \frac{2}{\pi} \int_0^\infty \r dk\, \frac{\sin(kx)}{k}\ \r e^{-Dk^2t}.
\end{equation*}
Therefore, the current at $x=0$ is 
\begin{equation*}
 j(t) = - \frac{2D}{\pi} \int_0^\infty \r dk\, \r e^{-Dk^2t} =
 - \frac{\sqrt D}{\sqrt{\pi t}}.
\end{equation*}
We recover hereby the usual exponent for diffusion.

The third example is a simple random walk in one dimension with the site
zero absorbing the walker. Let $p(x,t)$ denote the probability that the
walker reaches the origin for the first time at time $t$ starting out at
site $x$. We may assume that $x\in\Nl$. The probabilities $p$ are determined
by the recursion relation
\begin{itemize}
\item
 $p(0,0)=1$ and $p(0,t)=0$ for $t>0$
\item
 $p(x,t)=0$ whenever $x>t$
\item
 $p(x,t)=\frac{1}{2}(p(x-1,t-1)+p(x+1,t-1))$.
\end{itemize}
It can be checked that the solution is given by
\begin{equation*}
 p(x,t) = \left\{ 
 \begin{array}{ll} \frac{1}{2^t} \Bigl\{
 \binom{t-1}{(t-x)/2}-\binom{t-1}{(t-x-2)/2}\Bigr\}\quad
 &t-x \text{ even integer}\\
 0& \text{else}.
 \end{array}
 \right.
\end{equation*}
The current at time $t$ is then given by
\begin{equation*}
 J(t) = \sum_{x=1}^t p(x,t) = \frac{1}{2^t} \binom{t-1}{[(t-1)/2]} \sim
 \frac{1}{\sqrt{2\pi t}}.
\end{equation*}
In this formula, $[x]$ denotes the largest integer less or equal to $x$.

Finally, a random walk on $\Ir^3$ with a trapping set $A$ leads to a total 
current $J_A(t) = \r dN_A(t)/\r dt$ where $N_A(t)$ is the total number of
particles trapped by the set $A$ up to time $t$. It turns out that
\begin{equation*}
 J_A(t) \sim C(A) + 2 (2\pi)^{-3/2} C(A)^2 t^{-1/2},
\end{equation*}
where $C(A)$ is the capacity of the set $A$~\cite{spi}. Again, the
behaviour of the current returns the relevant information on the transport
properties of the system.

\section{Traps in Hilbert space}
\label{section2}

We first consider an abstract model of a trap that absorbs particles. An
explicit connection with a model of Fermions will be presented in
Section~\ref{section3}. The basic ingredients of our description are an
infinite dimensional Hilbert space $\g H$, a unitary operator $U$ on $\g H$
which specifies the evolution during a single time step, and a non-negative
operator $A$ less or equal than $\idty$ which describes the effect of the trap.
We shall assume that the trap is local in the sense that it
is of finite rank $d$. Writing its eigenvalue decomposition 
\begin{equation*}
 A = \sum_{j=1}^d p_j\, |\psi_j\>\<\psi_j|, \quad 0\le p_j\le 1
\end{equation*}   
we can think of $p_j$ as the probability that a particle in the state
$|\psi_j\>$ is absorbed by the trap when hitting the trap once. A value of
$p_j$ close to 1 means that the trap captures particles very efficiently in
the state $|\psi_j\>$. 

Starting out with a density $0\le\rho\le\idty$, the density after one time step
and hitting once the trap becomes 
\begin{equation*}
 (\idty-A) U\, \rho\, U^* (\idty-A) = (TU) \rho (TU)^*
\end{equation*} 
with $T := \idty - A$. Assuming a uniform initial distribution, i.e.\ $\rho$
a scalar multiple of $\idty$, the number
of particles absorbed up to time $t$ by the trap is proportional to 
\begin{equation}
 N_A(t) := \tr \Bigl( \idty - (T U)^t (U^*T)^t \Bigr) .
\label{1} 
\end{equation}  
The operator $(T U)^t (U^*T)^t$ is a finite rank perturbation of $\idty$,
therefore formula~(\ref{1}) makes sense. 

We shall always assume that eventually infinitely many particles are
absorbed by the trap. This will certainly be the case if the dynamics has
reasonable randomising properties. We are in particular interested in the
scaling behaviour of $N_A(t)$ with time, i.e., in the exponent $\gamma$
governing the asymptotics of $N_A$
\begin{equation*}
 N_A(t) \sim t^{1-\gamma}, \quad\text{$t$ large.} 
\end{equation*} 
The exponent $\gamma$ is non-negative as the growth of $N_A$ is at most
linear in time and cannot exceed the value~1 by our
assumption $\lim_{t\to\infty} N_A(t) = \infty$. We
expect this scaling to be related to the spectral properties of $U$ and to
be independent of the size of the trap. Using the telescopic formula
\begin{equation*}
 \idty - (T U)^t (U^*T)^t = \sum_{s=0}^{t-1} (T U)^s(\idty - T^2)(U^*T)^s,
\end{equation*}
we rewrite for $t\ge1$ $N_A(t)$ in terms of a current $J_A$
\begin{equation*}
 N_A(t) = \sum_{s=1}^t J_A(s)
\end{equation*} 
with
\begin{align}
 J_A(t) 
 &:= N_A(t)-N_A(t-1)
\nonumber \\ 
 &= \tr \Bigl\{ \bigl( \idty - (T U)^t (U^*T)^t \bigr) - \bigl( \idty - (T
 U)^{t-1} (U^*T)^{t-1} \bigr)\Bigr\}
\nonumber \\ 
 &= \tr (T U)^{t-1}(\idty - T^2)(U^*T)^{t-1}.
\label{2}
\end{align}
The expected scaling behaviour of $J_A$ is then
\begin{equation*}
 J_A(t) \sim t^{-\gamma}, \quad\text{$t$ large.} 
\end{equation*} 

The actual analysis will be performed for the simple case where $A$ is a
one-dimensional projector $|\varphi\>\<\varphi|$ with $\varphi$ a
normalised vector in $\g H$ and we shall henceforth drop the subscripts of
$N$ and $J$. In this case, $N$ is fully determined by the probability
measure
\begin{equation}
 \mu(\r d\theta) := \|E(\r d\theta)\varphi\|^2 
\label{3}
\end{equation}
on the unit circle $S^1$ where $E$ is the spectral measure of $U$
\begin{equation*}
 U = \int_{S^1} \r e^{-i \theta}\, E(\r d\theta).
\end{equation*} 
The expression~(\ref{2}) for the current now becomes
\begin{equation}
 J(t) = \|(\idty - P_{t-1}) \cdots (\idty - P_1)\, \varphi\|^2,\quad t>1
\end{equation}
with $P_t$ the orthogonal projector on $U^{-t} \varphi$. For consistency we
must put $J(1)=1$. As $(\idty-P_t)$ is a
contraction, $J$ is a monotonically decreasing function and therefore 
\begin{equation*}
 J_\infty := \lim_{s\to\infty} J(s)
\end{equation*}
exists. The behaviour of the current provides information on the transport
properties of the system. As we don't have in our general description a
notion  of position operator, allowing a definition of ballistic or
diffusive motion in terms of  spatial spreadings of wave functions, we
shall rather concentrate on the relation between the current and the
randomising properties of the dynamics which are quantified by the
dynamical entropy. This is, within the context of non-interacting Fermion
systems, the subject of Section~\ref{section3}. In this section we shall
study the asymptotic current in terms of the dynamical properties of the
trap states for the simple one-state trap. The analysis could be extended to
more involved traps that have a spatial structure. The more complicated
behaviour of the subsequent current could then be studied as in the case of
the 3D classical random walk, leading possibly to quantum capacities.   

In order to relate the measure $\mu$ with $J_\infty$, we introduce for $|z|<1$ 
the function
\begin{equation}
 G(z) := \sum_{s=1}^\infty z^s\, \mu^\wedge(s),
\label{6}
\end{equation}
where $\mu^\wedge$ is the Fourier transform of $\mu$
\begin{equation}
 \mu^\wedge(t) := \int_{S^1} \mu(\r d\theta)\, \r e^{-it\theta}, \quad
 t\in\Ir.
\label{7}
\end{equation}
Expressing $G$ in terms of $\mu$, we find
\begin{equation*}
 G(z) = \int_{S^1} \mu(\r d\theta)\, \frac{z}{\r e^{i\theta}-z}.
\end{equation*}
Obviously, $G$ is analytic in the open unit disc and we shall be concerned
with its value on the boundary of the disc. Writing $z = r\r e^{i\eta}$
with $0\le r<1$, a direct computation shows that 
\begin{equation}
 1 + 2\Re\g eG(r\r e^{i\eta}) = \int_{S^1} \mu(\r d\theta)\,
 \frac{1-r^2}{1+r^2- 2r\cos(\eta-\theta)}. 
\label{8} 
\end{equation}
The function 
\begin{equation*}
 \delta_r(\theta) := \frac{1-r^2}{1+r^2-2r\cos\theta},
\end{equation*}
is the Poisson kernel and the $\delta_r$ are a $\delta$-convergent
sequence of smooth, positive, normalised functions 
\begin{equation*}
 \frac{1}{2\pi} \int_0^{2\pi} \r d\theta\, \delta_r(\theta) = 1
\end{equation*} 
and 
\begin{equation}
 f(\eta) = \lim_{r\uparrow1} \frac{1}{2\pi} \int_{S^1} \r d\theta\,
 f(\theta)\  \delta_r(\theta-\eta)\quad \text{a.e.}
\label{15}
\end{equation}
for any integrable function $f$ on the unit circle. By a.e.\ we shall always
mean almost everywhere with respect to the Lebesgue measure.  

The imaginary part of $G$ is given by
\begin{equation*}
 \Im\g mG(r\r e^{i\eta}) = \int_{S^1} \mu(\r d\theta)\,
 \frac{\sin(\eta-\theta)}{1 + r^2 -2r\cos(\eta-\theta)}.
\end{equation*} 
When $r$ tends to 1, we obtain the Hilbert transform of $\mu$~\cite{coo,loo}
\begin{align}
 (\c H\mu)(\eta) 
 &= \lim_{r\uparrow1} \int_{S^1} \mu(\r d\theta)\, \frac{\sin(\eta-\theta)}{1
 + r^2 - 2r\cos(\eta-\theta)} 
\nonumber \\
 &= \lim_{\delta\downarrow0} \frac{1}{2} \int_{|\eta-\theta|\ge\delta}
 \mu(\r d\theta)\,\cot(\frac{\eta-\theta}{2}). 
\label{hilberttransform}
\end{align}
The limits in~(\ref{hilberttransform}) exist almost everywhere and for each
$\epsilon>0$ the set on which $|\c H\mu|$ is larger than $1/\epsilon$ has
Lebesgue measure less or equal to $\epsilon$. We shall now express the
current and asymptotic current in terms of $G$ and thus in terms of the
spectral properties of the trap, i.e.\ of the measure $\mu$.

\begin{lemma}
\label{lemma1}
 With the notation of above
 \begin{equation}
  J(t) = 1 - \sum_{s=1}^{t-1} |K(s)|^2, 
 \label{lemma11} 
 \end{equation}
 where the function $K$ is determined by the relation
 \begin{equation}
  F(z) := \frac{G(z)}{1+G(z)} = \sum_{s=1}^\infty z^s\, K(s), \quad |z|<1. 
 \label{lemma12}
 \end{equation}
 The asymptotic current is given by
 \begin{equation}
  J_\infty = \lim_{r\uparrow1} \frac{1}{2\pi} \int_{S^1} \r d\theta\,
  \frac{1 + 2\Re\g eG(r\r e^{i\theta})}{|1+G(r\r e^{i\theta})|^2}. 
 \label{lemma13}
 \end{equation}
\end{lemma}

\begin{proof}
We introduce
\begin{equation*}
 K(t) := \< (U^*)^t \varphi, (\idty - P_{t-1}) \cdots (\idty - P_1)\, \varphi\>
\end{equation*}
and denote by $F$ the $Z$-transform of $K$
\begin{equation}
 F(z) := \sum_{s=1}^\infty z^s\, K(s), \quad |z|<1. 
\label{11} 
\end{equation}

The relation~(\ref{lemma11}) follows from a straightforward computation
\begin{align*}
 J(t)
 &= \|(\idty - P_{t-1}) \cdots (\idty - P_1)\, \varphi\|^2 \\
 &= \|(\idty - P_{t-2}) \cdots (\idty - P_1)\, \varphi\|^2 - |\< (U^*)^{t-1}
 \varphi, (\idty - P_{t-2}) \cdots (\idty - P_1)\, \varphi\>|^2 \\
 &= J(t-1) - |K(t-1)|^2 .
\end{align*}

Next, we write
\begin{equation*}
 (\idty - P_t) \cdots (\idty - P_1)\, \varphi = 
 (\idty - P_{t-1}) \cdots (\idty - P_1)\, \varphi - K(t)\, (U^*)^t \varphi. 
\end{equation*}
Therefore
\begin{equation}
 (\idty - P_t) \cdots (\idty - P_1)\, \varphi = \varphi - \sum_{s=1}^t K(s)\,
 (U^*)^s  \varphi.
\label{4} 
\end{equation}
Taking the scalar product of $(U^*)^{t+1}\varphi$ with~(\ref{4}), we obtain
\begin{equation}
 K(t+1) = \< (U^*)^{t+1} \varphi, \varphi\> - \sum_{s=1}^t \< (U^*)^{t-s+1}
 \varphi,  \varphi\>\, K(s).
\label{5}
\end{equation}
Equation~(\ref{5}) has the structure of a one-sided convolution equation. Using
the $Z$-transform it becomes an algebraic equation. More precisely, 
multiplying~(\ref{5}) with $z^{t+1}$ and summing from $t=0$ to $\infty$, we
obtain
\begin{equation*}
 F(z) = G(z) - F(z)\, G(z). 
\end{equation*}  
As by~(\ref{8}) $1 + G$ never vanishes inside the unit disk
\begin{equation*}
 \frac{G(z)}{1+G(z)} = F(z) = \sum_{s=1}^\infty z^s\, K(s),
\end{equation*}
proving~(\ref{lemma12}).

The basic relation between the asymptotic current $J_\infty$ and the measure 
$\mu$ is obtained by
applying Parseval's formula to~(\ref{11}). Putting $z=r\r e^{i\eta}$
\begin{equation}
  \frac{1}{2\pi} \int_{S^1} \r d\eta\, \Bigl| F(r \r e^{i\eta}) \Bigr|^2 = 
  \sum_{s=1}^\infty r^{2s}\, |K(s)|^2. 
\label{12}
\end{equation}
We compute now the asymptotic current on the basis of~(\ref{lemma11}) 
\begin{align}
 J_\infty 
 &= \lim_{t\to\infty} J(t)
 = 1 - \lim_{t\to\infty} \sum_{s=1}^{t-1} |K(s)|^2 
\nonumber \\
 &= 1- \lim_{r\uparrow1} \sum_{s=1}^\infty r^{2s}\, |K(s)|^2 
 = 1- \lim_{r\uparrow1} \frac{1}{2\pi} \int_{S^1} \r d\theta\, |F(r\r
 e^{i\theta})|^2 
\label{20} \\
 &= \lim_{r\uparrow1} \frac{1}{2\pi} \int_{S^1} \r d\theta\,
  \frac{1 + 2\Re\g eG(r\r e^{i\theta})}{|1+G(r\r e^{i\theta})|^2}.
\nonumber
\end{align} 
\end{proof}

Our first result deals with the asymptotic current for trap states with 
absolutely continuous dynamical spectrum. 

\begin{theorem}
\label{theorem1}
 Suppose that $\mu$ is absolutely continuous w.r.t.\ the Lebesgue measure,
 then $J_\infty>0$.
\end{theorem}

\begin{proof}
Let $\mu$ be absolutely continuous with respect to the Lebesgue measure with 
density $\rho$. We have for $r<1$ 
\begin{equation*}
 (1+2\Re\g eG)(r\r e^{i\eta}) =  \frac{1}{2\pi} \int_{\c S^1} \r d\theta\,  
 \rho(\theta+\eta)\, \frac{1-r^2}{1-2r \cos\theta+r^2}.
\end{equation*} 
Because $\rho$ is integrable and because of the properties of the Poisson
kernel
\begin{equation*}
 \lim_{r\uparrow1} (1+2\Re\g e G)(r \r e^{i\eta}) =  \rho(\eta)\quad
 \text{a.e.}
\end{equation*} 
Also the imaginary part of 
\begin{equation*}
 z \mapsto \frac{1}{2\pi} \int_{\c S^1} \r d\theta\, \rho(\theta)\, 
 \frac{z}{\r e^{i\theta} - z}
\end{equation*} 
converges almost everywhere to the Hilbert transform $\c H\rho$ of $\rho(\theta)\r d\theta$:
\begin{equation*}
 \lim_{r\uparrow1}\Im\g mG(r\r e^{i\eta}) = (H\rho)(\eta) := \lim_{\delta\downarrow0}          \frac{1}{4\pi} \int_{|\theta|\ge\delta} \r d\theta\, 
 \rho(\theta+\eta)\, \cot(\theta/2).
\end{equation*}
In the expression for the asymptotic current 
\begin{align*}
 J_\infty 
 &= \lim_{r\uparrow1} \frac{1}{2\pi} \int_{\c S^1} \r d\theta\, 
 (1 - |F(r\r e^{i\theta})|^2) \\
 &= \lim_{r\uparrow1} \frac{1}{2\pi} \int_{\c S^1} \r d\theta\, 
 \frac{1+2\Re\g eG(r\r e^{i\theta})}
 {1 + 2\Re\g eG(r\r e^{i\theta}) + (\Re\g eG)^2(r\r e^{i\theta}) +
 (\Im\g m G)^2(r\r e^{i\theta})} 
\end{align*}  
the integrand is bounded by 1 and tends almost everywhere to
\begin{equation*}
 \frac{4\rho(\theta)}{(1 + \rho(\theta))^2 + 4(\c H\rho)^2(\theta)}
\end{equation*}
as $r$ grows to 1. We can therefore apply the dominated convergence theorem 
to obtain
\begin{equation*}
 J_\infty = \frac{1}{2\pi} \int_{\c S^1} \r d\theta\, 
 \frac{4\rho(\theta)}{(1 + \rho(\theta))^2 + 4(\c H\rho)^2(\theta)} > 0.
\end{equation*} 
\end{proof}

In~(\ref{20}), we have replaced a limit $t\to\infty$ by a limit
$r\uparrow1$. In fact more information can be gained in doing so. Indeed, in
the case $J_\infty=0$, the behaviour of $J(t)$ for large $t$ is that of
$1- \sum_{s=1}^{t-1} |K(s)|^2$. The function
\begin{equation}
 \tilde J(r) := 1- \sum_{s=1}^\infty r^{2s}\, |K(s)|^2,\qquad 0\le r<1
\label{21}
\end{equation}
is expressed in terms of the discrete Laplace transform of $s\mapsto |K(s)|^2$
and Tauberian theorems relate the large time behaviour of $t\mapsto J(t) = 
1- \sum_{s=1}^{t-1} |K(s)|^2$ with that of $\tilde J(r)$ when $r\uparrow1$.
We shall first use this idea to show that $J_\infty$ vanishes when $\mu$ is
singular. Next, we shall in a few examples consider convergence exponents.   
 
\begin{theorem}
\label{theorem2}
 If $\mu$ is singular w.r.t.\ the Lebesgue measure, then $J_\infty=0$. 
\end{theorem}

\begin{proof}
 Let $\mu$ be singular and hence concentrated on a measurable set with zero
 Lebesgue measure. As the Lebesgue measure of this set equals the infimum
 of the Lebesgue measures of open subsets containing the set, we can given
 any positive $\epsilon$ find an open subset $A$ of $\c S^1$  such that the
 Lebesgue measure of $A$ is not larger than $\epsilon$ and $\mu(A)=1$. The
 set $A$ is a countable union of disjoint open intervals
 $]\alpha_j,\beta_j[$.  We dress each of the $]\alpha_j,\beta_j[$ with open
 strips of width $\delta_j$ which shall be determined later on. By choosing
 the $\delta_j$ sufficiently small we may still ensure that the Lebesgue
 measure of $A^+ := \bigcup_j ]\alpha_j-\delta_j, \beta_j+\delta_j[$ is
 small.  
 
We shall also need   
\begin{align}
 \frac{1}{2\pi} \int_{|\theta| \ge \delta} \r d\theta\,
 \frac{1-r^2}{1+r^2-2r\cos\theta} 
 &\le 1 \quad\text{for } \delta \le 1-r 
 \nonumber \\
 &\le \frac{1-r}{\delta} \quad\text{for } \delta \ge 1-r. 
 \label{tailpoisson}
\end{align}
The estimate for the case $\delta>1-r$ is obtained as follows: 
\begin{align*}
 \frac{1}{2\pi} \int_{|\theta|\ge\delta} \r d\theta\,
 \frac{1-r^2}{1+r^2-2r\cos\theta}
 &= \frac{1}{\pi} \int_\delta^\pi \r d\theta\,
 \frac{1-r^2}{1+r^2-2r\cos\theta} \\
 &\le \frac{2(1-r)}{\pi} \int_\delta^\pi \r d\theta\,
 \frac{1}{1+r^2-2r\cos\theta} \\
 &= \frac{2(1-r)}{\pi} \int_\delta^\pi \r d\theta\,
 \frac{1}{(1-r)^2 + 2r(1-\cos\theta)} \\
 &\le \frac{(1-r)}{\pi r} \int_\delta^\pi \r d\theta\,
 \frac{1}{1-\cos\theta} \\ 
 &\le \frac{1-r}{\delta} \quad \text{for $\delta$ sufficiently small.}
\end{align*}
 
We now estimate the current
\begin{align}
 \tilde J(r)
 &= \frac{1}{2\pi} \int_{\c S^1} \r d\eta\, \Bigl( 1-|F(r\r e^{i\eta})|^2\Bigr)
 \nonumber \\
 &= \frac{1}{2\pi} \int_{A^+} \r d\eta\,  \Bigl( 1-|F(r\r
 e^{i\eta})|^2\Bigr) + \frac{1}{2\pi} \int_{\c S^1 \setminus A^+} \r
 d\eta\,  \Bigl( 1-|F(r\r e^{i\eta})|^2\Bigr)
 \nonumber \\
 &\le \epsilon + 2\sum_j \delta_j + \frac{1}{2\pi} \int_{\c S^1 \setminus
 A^+} \r d\eta\, \Bigl( 1-|F(r\r e^{i\eta})|^2\Bigr)
 \nonumber \\
 &= \epsilon + 2\sum_j \delta_j + \frac{1}{2\pi} \int_{\c
 S^1 \setminus A^+} \r d\eta\, \frac{1+2\Re\g eG}{1
 +  2\Re\g eG+(\Re\g eG)^2+ (\Im\g  mG)^2}(r\r e^{i\eta}) 
 \nonumber \\
 &\le \epsilon +2 \sum_j \delta_j + \frac{1}{2\pi} 
 \int_{\c S^1 \setminus A^+} \r d\eta\, 
 \frac{1+2\Re\g eG}{1 + 2\Re\g eG+(\Re\g eG)^2} (r\r e^{i\eta}) 
 \nonumber \\
 &\le \epsilon + 2\sum_j \delta_j + \frac{2}{\pi} 
 \int_{\c S^1 \setminus A^+} \r d\eta\, \int_{\c S^1}
 \mu(\r d\theta)\, \frac{1-r^2}{1+r^2-2r\cos(\eta-\theta)}
 \nonumber \\
 &= \epsilon + 2\sum_j \delta_j + \frac{2}{\pi} 
 \int_{\c S^1 \setminus A^+} \r d\eta\, \int_{\c S^1}
 \mu(\r d\theta)\, \delta_r(\eta-\theta)
 \nonumber \\
 &\le \epsilon + 2\sum_j \delta_j + \frac{2}{\pi} \int_{A} \mu(\r
 d\theta)\, \int_{\c S^1 \setminus A^+} \r d\eta\, \delta_r(\eta-\theta).
 \label{estimate1}
\end{align}
In the last but one inequality we have used that the integrand is bounded from above
by $4(1+2\Re\g eG)$. In the last inequality we use~(\ref{tailpoisson}) to get 
\begin{align}
 \tilde J(r) \le 
 &\epsilon + 2\sum_j \delta_j + \frac{2}{\pi}\Bigl( \sum_{\substack{k \\
 \delta_k < 1-r}} \mu(]\alpha_k,\beta_k[)\Bigr) 
\nonumber \\
 &+ \frac{2}{\pi}(1-r)\Bigl( \sum_{\substack{k \\ \delta_k \ge  1-r}}
 \frac{\mu(]\alpha_k,\beta_k[)}{\delta_k} \Bigr). 
\label{estimate3} 
\end{align}
For a given $\epsilon$ and a given large $N$, we may take $\delta_k =
N \mu(]\alpha_k,\beta_k[) (1-r)$. When $r$ is sufficiently close to 1, the upper
bound for $J$ becomes
\begin{align}
 \tilde J(r) \le 
 &\epsilon + 2N(1-r) + \frac{2}{\pi} \Bigl(\sum_{\substack{k \\
 N\mu(]\alpha_k,\beta_k[) < 1}} \mu(]\alpha_k,\beta_k[) \Bigr) 
\nonumber \\ 
 &+ \frac{2}{N\pi}
 \#\Bigl( \{k \mid N\mu(]\alpha_k,\beta_k[)\ge 1\}\Bigr). 
\label{estimate4}
\end{align}
First we fix an arbitrary small $\epsilon$ and the corresponding sets
$]\alpha_k,\beta_k[$. The last two terms in~(\ref{estimate4}) can be made
small by choosing $N$ sufficiently large. Finally the second term
in~(\ref{estimate4}) becomes small when we let $r\uparrow1$.    
\end{proof}

We conclude this section with some examples, remarks and partial results
about exponents. Let us assume that we are in the situation $J_\infty=0$ and
that we can assign an exponent to $J$, i.e.
\begin{equation*}
 \gamma := \lim_{t\to\infty} - \frac{\log J(t)}{\log t}
\end{equation*}
exists. We shall, moreover, assume that infinitely many particles
eventually are absorbed by the trap and actually strengthen this condition
to $\gamma<1$.  We are interested in relating the behaviour of $t\mapsto
J(t)$ as $t\to\infty$  with that of $r\mapsto\tilde J(r)$ as $r\uparrow1$.
We therefore introduce upper and lower exponents $\overline\alpha$ and
$\underline\alpha$ for $\tilde J$
\begin{equation*}
 \overline \alpha := \limsup_{r\uparrow1} \frac{\log \tilde J(r)}{\log(1-r)}
 \qquad\text{and}\qquad
 \underline \alpha := \liminf_{r\uparrow1} \frac{\log \tilde J(r)}{\log(1-r)}.
\end{equation*} 

\begin{lemma}
 Using the notations and assumptions of above, $\overline\alpha =
 \underline\alpha = \gamma$.
\end{lemma}

\begin{proof}
 We have to consider
 \begin{align*}
  &J(t) = 1 - \sum_{s=1}^{t-1} |K(s)|^2 = \sum_{s=t}^\infty |K(s)|^2
  \qquad\text{and} \\ 
  &\tilde J(r) = 1 - \sum_{s=1}^\infty r^{2s}\, |K(s)|^2 = \sum_{s=1}^\infty
  (1-r^{2s}) |K(s)|^2.
 \end{align*} 
For notational convenience, we put $c_t := |K(t)|^2$ and $\lambda := -2 \log
r$. 

Let $\alpha>\gamma$, then
\begin{align*}
 \lim_{\lambda\downarrow0} \lambda^{-\alpha} \sum_{s=1}^\infty (1-\r
 e^{-\lambda s}) c_s 
 &\ge \lim_{\lambda\downarrow0} (1-\frac{1}{\r e})\, 
 \lambda^{-\alpha} \sum_{s=[\lambda^{-1}]}^\infty c_s \\
 &= \lim_{N\to\infty}  (1-\frac{1}{\r e})\,  N^\alpha \sum_{s=N}^\infty c_s
 = \infty.
 \end{align*}
Hence, $\underline\alpha\ge\gamma$. 

Conversely, let $\alpha<\gamma$ and introduce the short notation $f(N) :=
\sum_{s=N}^\infty c_s$. Fix an arbitrary $\epsilon>0$ and a $\gamma_0$ such
that $\alpha<\gamma_0<\gamma$. For $\lambda>0$ and $\kappa=2,3,\ldots$, let
$\tilde N_n$ be determined by
\begin{equation*}
 \exp\bigl(-\lambda \tilde N_n\bigr) = \frac{\kappa - n}{\kappa},\quad 
 n=1,2,\ldots,\kappa -1 
\end{equation*}
and put $N_n := [\tilde N_n]^+$ where $[x]^+$ is the smallest integer larger
or equal to $x$. We shall pick $\kappa$ later on in such a way
that $\kappa\ll\lambda^{-1}$. This implies that the $N_j$ are far apart, in
particular that $N_1\gg1$. We now have obtained a partition
\begin{equation*} 
 1 =: N_0 \ll N_1 \ll \cdots \ll N_{\kappa-1} \ll N_{\kappa} := +\infty 
\end{equation*} of $\Nl_0$ such that
\begin{equation*}
 1-\r e^{-\lambda t} \le \frac{n+1}{\kappa}
 \quad\text{for } N_n\le t< N_{n+1}.
\end{equation*}
\begin{align}
 \sum_{s=1}^\infty (1-\r e^{-\lambda s}) c_s 
 &= \sum_{n=0}^{\kappa-1} \sum_{s=N_n}^{N_{n+1}-1} (1-\r e^{-\lambda s})
 c_s
\nonumber \\ 
 &\le \sum_{n=0}^{\kappa-1} \frac{n+1}{\kappa} \sum_{s=N_n}^{N_{n+1}-1}
 c_s
\nonumber \\ 
 &= \sum_{n=0}^{\kappa-1} \frac{n+1}{\kappa} \bigl(f(N_n) - f(N_{n+1})\bigr)
\nonumber \\
 &\le \frac{1}{\kappa} \Bigl( f(N_0) + f(N_1) + \cdots + f(N_{\kappa-1})
 \Bigr)
\nonumber \\
 &\le \frac{1}{\kappa} \Bigl(1 + \epsilon N_1^{-\gamma_0} +\cdots+ \epsilon 
 N_{\kappa-1}^{-\gamma_0}\}.
\label{22} 
\end{align}
The inequality in~(\ref{22}) holds for $N_1$ large enough, i.e.\ for
$\kappa$ sufficiently large.
Because $\tilde N_n = [\frac{1}{\lambda} \log(\frac{\kappa}{\kappa-n})]^+$
we have
\begin{equation*} 
 N_n^{-\gamma_0} < \left( \frac{1}{\lambda} \log\left(
 \frac{1}{1-\frac{n}{\kappa}} \right)  \right)^{-\gamma_0}
\end{equation*}
and thus
\begin{align*}
 \sum_{s=1}^\infty (1-\r e^{-\lambda s}) c_s 
 &\le \frac{1}{\kappa} + \frac{1}{\kappa} \epsilon \sum_{n=1}^{\kappa-1}
 \left( \frac{1}{\lambda} \log\left(\frac{1}{1-\frac{n}{\kappa}} \right)
 \right)^{-\gamma_0} \\
 &\le \frac{1}{\kappa} + \lambda^{\gamma_0} \epsilon \frac{1}{\kappa}
 \sum_{n=1}^{\kappa-1} \Bigl(  \log\bigl(\frac{1}{1-\frac{n}{\kappa}}\bigr)
 \Bigr)^{-\gamma_0} \\
 &\le \frac{1}{\kappa} + \lambda^{\gamma_0} \epsilon \int_0^1 \r dy\, \Bigl(
 \log\bigl( \frac{1}{1-y}\bigr) \Bigr)^{-\gamma_0} \\
 &= \frac{1}{\kappa} + \lambda^{\gamma_0} \epsilon \int_0^1\r dt\, \Bigl(
 \log \frac{1}{t} \Bigr)^{-\gamma_0} \\
 &= \frac{1}{\kappa} + \lambda^{\gamma_0} \epsilon \Gamma(1-\gamma_0) \\
 &\le \delta \lambda^{\gamma_0}
\end{align*}
with $\delta$ arbitrarily small. This last inequality is obtained by choosing
$\lambda^{-\gamma_0} \ll \kappa \ll \lambda^{-1}$ which is possible because
$0\le\gamma_0<1$. Hence $\overline\alpha\le\gamma$ and the lemma is proven.
\end{proof}

When estimating the current in the proof of
Theorem~\ref{theorem2} we dropped the contribution of $\Im\g mG$. Generally,
this may lead to underestimate the exponent. A simple example is provided by
a measure $\mu$ that is concentrated on a finite set such as $\mu(\r
d\theta) = \delta(\theta)\r d\theta$ (the general case being quite
similar). A simple calculation shows that $F(z)=z$.
Therefore 
\begin{equation*}
 \frac{1}{2\pi} \int_{\c S^1} \r d\eta\, \Bigl( 1-|F(r\r
 e^{i\eta})|^2\Bigr) = 1-r^2  \sim  2(1-r)
\end{equation*} 
and the true exponent is 1. Dropping the imaginary part of $G$ in the
integral, we get an exponent 1/2:
\begin{align*}
 \frac{1}{2\pi} \int_{\c S^1} \r d\eta\, \frac{1+2\Re\g eG}{1+2\Re\g
 eG+(\Re\g eG)^2} (r\r e^{i\eta})
 &= \frac{1}{2\pi} \int_{\c S^1} \r d\eta\,
 \frac{(1-r^2)(1+r^2-2r\cos\eta)}{(1-r\cos\eta)^2} \\
 &\sim \int \r dx\, \frac{(1-r) x^2}{(2(1-r)+x^2)^2} \\
 &\sim \sqrt{1-r}.
\end{align*} 
  
Let $\mu$ be a discrete measure, possibly concentrated on a dense subset of
$\c S^1$, 
\begin{equation*}
 \mu(\r d\theta) = \sum_j \rho_j \delta(\theta-\theta_j) \r d\theta
\end{equation*}
with
\begin{equation*}
 \rho_j\ge0, \quad \sum_j \rho_j=1 \quad \text{and } i\ne j\Rightarrow
 \theta_i\ne\theta_j.
\end{equation*}
Applying the estimates in the proof of Theorem~\ref{theorem2} we obtain
\begin{equation}
 \tilde J(r) \le 2\sum_j \delta_j + 4 \sum_{\substack{j \\ \delta_j\ge 1-r}} 
 \frac{\rho_j(1-r)}{\delta_j} + 4 \sum_{\substack{j \\ \delta_j\le 1-r}} \rho_j.
\label{currentatomic}
\end{equation}

Suppose that the $\rho_j$ tend sufficiently rapidly to zero in order that
also $\sum_j \sqrt{\rho_j}<\infty$. Choosing $\delta_j=\sqrt{(1-r)\rho_j}$,
we obtain $\tilde J(r) \le \bigl(\sum_j \sqrt{\rho_j}\bigr) \sqrt{1-r}$ and
therefore an exponent 1/2. In view of the first example, this is the best
exponent we can hope to obtain neglecting the contribution of the imaginary
part of $G$. 

Suppose that $\rho_j \sim j^{-\alpha}$ with $\alpha>1$. Choosing
in~(\ref{currentatomic}) $\delta_j = N\rho_j(1-r)$ with $N$ large, we obtain
\begin{equation*}
 \tilde J(r) \le 2N(1-r) + \frac{2}{N\pi} \#\Bigl( \{j \mid \rho_j\ge
 \frac{1}{N}\}\Bigr) + \frac{2}{\pi} \sum_{\substack{j \\ \rho_j\le 
 \frac{1}{N}}} \rho_j.
\end{equation*}
The optimal choice for $N$ is $N \sim (1-r)^{-\alpha/(2\alpha-1)}$ and this 
yields an exponent $(\alpha-1)/(2\alpha-1) < 1/2$. The exponent $1/2$ is 
reached for all $\rho_j$ that tend faster to zero that any inverse power of $j$.
 
In our last example the measure $\mu$ will be singular continuous~\cite{sim}. 
Given a number $x\in [0,1]$, we write its binary expansion as
\begin{equation*}
 x = \sum^{\infty}_{m=1} \frac{a_m(x)}{2^m},
\end{equation*}
where $a_m(x)\in \{0,1\}$ for $m \ge 1$.
This defines a map $F:\{0,1\}^{\Nl}\to [0,1]$ which can be used to
transport a measure $\lambda$ on $\{0,1\}^{\Nl}$ to a measure $\mu$ on
$[0,1]$ by the relation $\mu(A)=\lambda(F^{-1}[A])$. Take for $\lambda$ the
infinite product of the measure $(1-p,p)$, $p\in [0,1]$. The corresponding
measure $\mu$ on $[0,1]$ is called the Bernoulli measure $\mu_p$. Except
for $p=0$, $p=1$ (Dirac measures) and $p=1/2$ (Lebesgue measure), $\mu_p$
is singular continuous.

We want to estimate the exponent $\alpha$ of the current $\tilde J(r)\sim
(1-r)^\alpha$. Assume $0<p<1/2$ and let $q$ be a number between $p$ and
$1/2$. Define the sets $A(n)$ for $n\ge1$ as
\begin{equation*}
 A(n) = \bigl\{ x\in [0,1] \,\bigm|\, \#\{m\le n \,|\, a_m(x)=1\} < qn \bigr\}.
\end{equation*}
The asymptotic behaviour of the Bernoulli measure is
\begin{equation*}
 1-\mu_p(A(n)) \sim \exp(-n S(q|p)),
\end{equation*}
whereas for the Lebesgue measure
\begin{equation*}
 |A(n)| \sim \exp(-n S(q|1/2)),
\end{equation*}
with $S(p_1|p_2)$ the relative entropy of the probability
measures $(p_1,1-p_1)$ with respect to $(p_2,1-p_2)$, i.e.,
\begin{equation*}
 S(p_1|p_2) = p_1\log(p_1) + (1-p_1)\log(1-p_1) - p_1\log(p_2) -
 (1-p_1)\log(1-p_2).
\end{equation*}
The sets $A(n)$ are finite unions of $K(n)$ intervals $A_k(n)$
for which the first $n$ digits of the binary expansion are given.
We have the asymptotics
\begin{equation*}
 K(n) \sim \exp(n S(q)),
\end{equation*}
where $S(q)$ is the (Shannon) entropy of the probability measure
$(q,1-q)$, i.e.,
\begin{equation*}
 S(q) := - q\log(q) - (1-q)\log(1-q).
\end{equation*}

As in the proof of Theorem~\ref{theorem2} we dress the intervals
$A_k(n)$ by small strips of length $\delta_k(n)$. This results in
the set $A^+(n)$. We can then use the estimate~(\ref{estimate1})
for the current $\tilde J(r)$
\begin{align*}
 &\int_{S\setminus A^+(n)} \r d\eta
 \left\{
   \int_{A(n)} \mu(\r d\theta) \delta_r(\eta-\theta)
 + \int_{S\setminus A(n)} \mu(\r d\theta) \delta_r(\eta-\theta)
 \right\} \\
 &\le \frac{2}{\pi} \sum_{\substack{k \\ \delta_k(n)<1-r}} \mu(A_k(n))
 + (1-r) \sum_{\substack{k \\ \delta_k(n)>1-r}} \frac{\mu(A_k(n))}{\delta_k}
 + (1-\mu_p(A(n))).
\end{align*}
Again taking $\delta_k(n)=N\mu(A_k(n))(1-r)$, we obtain
\begin{align*}
 \tilde J(r) \le 
 &|A(n)| + 2N(1-r)
 + \sum_{\substack{k \\ N\mu(A_k(n))<1}} \mu(A_k(n)) \\
 &+ \frac{1}{N} \#\{k\,|\,N\mu(A_k(n))>1\} + (1-\mu_p(A(n))).
\end{align*}

Imposing now the scaling behaviour
$N \sim \exp(n\nu)$ and $1-r \sim \exp(-n\beta)$,
we find
\begin{equation*}
 \alpha \ge \left\{
 \frac{\min\{S(q|1/2),\beta-\nu,\nu-S(q),S(q|p)\}}{\beta}
 \right\}
\end{equation*}
for any $q$, $\nu$ and $\beta$. The optimal values are
\begin{equation*}
 q = \frac{\log 2(1-p)}{\log\frac{p}{1-p}}
\end{equation*}
(for which $S(q|1/2)=S(q|p)$), 
$\nu=\log 2$ and $\beta=2\log 2 - S(q)$.
Finally, the lower bound for the exponent $\alpha$ is
\begin{equation}
\label{boundexp}
 \alpha \ge \frac{\log 2 - S(q)}{2\log 2 - S(q)}.
\end{equation}

\begin{figure}
\begin{center}
\includegraphics[width=13cm]{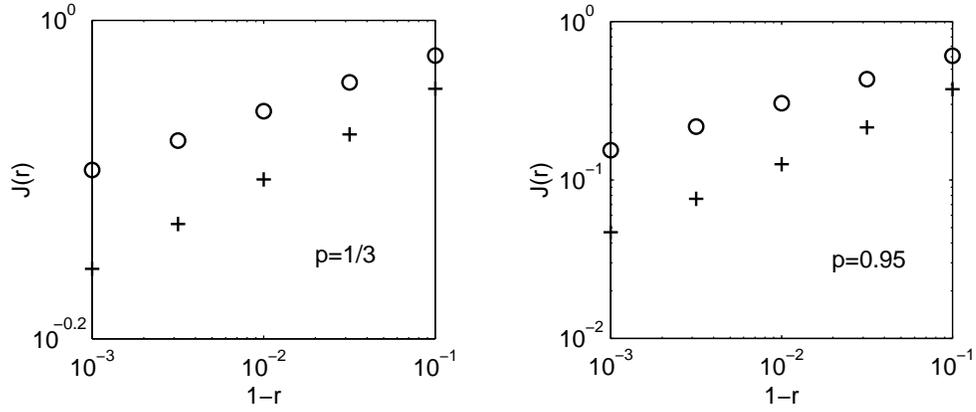}
\end{center}
\caption{\label{figure}
The current $\tilde J(r)$, with (crosses) and without (circles)
the contribution $\Im\g mG$. Left, $p=1/3$ and right, $p=0.95$.}
\end{figure}

We compare this bound with some numerical computations.
For a few pairs $(p,r)$ the integrals for $G(z)$ and $\tilde J(r)$
were evaluated on an equidistant mesh of $2^{13}$ points.
To illustrate the approximation made in (\ref{estimate1})
by dropping $\Im\g mG$, we performed the computation
both with and without this imaginary part.
The relative precision for the current $\tilde J(r)$ was checked to be
better than $0.01$, while for the exponent $\alpha$ it is $0.1$.
Figure~\ref{figure} shows the existence of the exponents and
the error introduced by neglecting the imaginary part of $G(z)$.
The latter is quantified in Table~\ref{table}. For this example
the lower bound (\ref{boundexp}) for $\alpha$ is rather sharp.

\begin{table}
\begin{center}
\begin{tabular}{c|ccc}
 & analytical & numerical & numerical \\
 & & without $\Im\g mG$ & with $\Im\g mG$ \\
\hline
$p=1/3$  & 2.05 10$^{-2}$ & 3.7 10$^{-2}$ & 5.6 10$^{-2}$ \\
$p=0.95$ & 1.96 10$^{-1}$ & 3.2 10$^{-1}$ & 4.2 10$^{-1}$ \\
\end{tabular}
\end{center}
\caption{\label{table}
Estimates for the exponent $\alpha$.}
\end{table}

\section{Dynamical entropy of a Fermion dynamics}
\label{section3}

We shall in this section apply our results in the setting of a 
free Fermionic gas. As this is a system of non-interacting particles,
it is completely described in terms of single-particle quantities. 
Second quantisation allows to lift one-particle objects to the 
many particles, taking into account the Fermi statistics. We remind here
briefly the mathematical setup.

We shall denote the single-particle Hilbert space by $\g H$. The
observables of the Fermion algebra, also called CAR for canonical
anticommutation relations, is the C*-algebra $\c A(\g H)$ determined
through the relations
\begin{equation*}
 a(f+\alpha g) = a(f)+ \overline\alpha a(g),
 \quad \{a(f),a(g)\}=0
 \quad\text{and}\quad
 \{a(f),a^*(g)\}=\<f,g\>.
\end{equation*}    
Sometimes we shall deal with a one-particle space $\g H$ of finite
dimension $d$. In this case $\c A(\g H)$ is easily seen to be isomorphic to
the algebra of matrices of dimension $2^d$. An explicit construction is
given in terms of linear transformations of the antisymmetric Fock space
$\Gamma(\g H)$ which is spanned by the $n$-particle vectors
\begin{equation*}
 a^*(f_1)a^*(f_2)\cdots a^*(f_n)\Omega
\end{equation*} 
for $0\le n\le d$. The normalised vector $\Omega$ is called the vacuum and
it is annihilated by any operator $a(f)$.

The construction of dynamical entropy presented in~\cite{alf,af} is based
on the following idea. Given a unital C*-algebra $\g A$ and a reference
state $\omega$, one considers an operational partition, i.e.\ a finite
collection $\g X = \{x_1,x_2,\ldots,x_n\}$ of elements of $\g A$ satisfying
$\sum_j x^*_jx_j = \idty$. This yields a correlation matrix 
\begin{equation*}
 \rho[\g X] := [\omega(x_j^*x_i)]
\end{equation*}  
with corresponding von~Neumann entropy
\begin{equation*}
 \En[\omega,\g X] := \tr \eta(\rho[\g X]),
\end{equation*}
where $\eta$ is the usual entropy function 
\begin{equation*}
 \eta(0):=0
 \qquad\text{and}\qquad
 \eta(x):=-x\log x, \quad 0<x\le1.
\end{equation*}
The average entropy of an operational partition $\g X = (x_1,x_2,\ldots,x_n
)$ arises by computing the average entropy of the correlation matrix $\g
X_t$ corresponding to the refinement of the partition $\g X$ at discrete
times up to $t-1$. More precisely
\begin{equation*}
 \g X_t := \Theta^{t-1}(\g X) \circ\cdots\circ \Theta(\g X) \circ \g X.
\end{equation*}
In this expression, 
\begin{align*}
 &\Theta^s(\g X) := \bigl(\Theta^s(x_1),\Theta^s(x_2),\ldots,\Theta^s(x_n)\bigr)
 \qquad\text{and} \\[4pt]
 &\g X\circ\g Y := (x_1y_1,x_2y_1,\ldots,x_ny_m)
 \quad\text{for } \g Y = (y_1,y_2,\ldots,y_m).
\end{align*}
It can happen that the growth of $t\mapsto \En[\omega,\g X_t]$ is sublinear
if the dynamics is not sufficiently randomising. In such a case one can look
for a growth exponent. 

We shall in the following pages obtain a lower bound
for $\En[\omega,\g X_t]$ in terms of particle numbers absorbed by a trap for
the case of a weakly interacting Fermion system, meaning that we may use an
effective one-particle dynamics $\Theta(a(f)) := a(Uf)$ for the evolution.
$U$ is a unitary operator on the one-particle space $\g H$.   

The reference state will be chosen accordingly as a gauge-invariant
quasi-free state. Such a state $\omega_Q$ is uniquely determined by its
symbol $Q$ which is a linear operator on $\g H$ satisfying $0\le Q\le
\idty$. The only monomials in the creation and annihilation operators $a^*$
and $a$ which have non-zero expectations contain a same number of each and
\begin{equation*}
 \omega_Q(a^*(f_1)\cdots a^*(f_n)a(g_n)\cdots a(g_1)) = \det(\<g_i,Q\,f_j\>).
\end{equation*} 
In particular, we may choose $Q = \kappa \idty$ for $0\le\kappa\le1$. Such
states are homogeneous and in e.g.\ the case of Fermions on a lattice
they describe independent Fermions occupying each site of the lattice
with probability $\kappa$. The Fock vacuum, i.e.\ the vector state
determined by $\Omega$ in the Fock representation of above, corresponds to
the choice $\kappa=0$. The choice $\kappa=1/2$ corresponds to the unique
tracial state on $\c A(\g H)$. A quasi-free state $\omega_Q$ is known to be
pure if and only if $Q$ is an orthogonal projector. Moreover, any
$\omega_Q$ can be obtained as the restriction of a pure quasi-free state on
a larger CAR algebra by using the purification construction. One introduces
the auxiliary space $\g K := \overline{Q(\idty-Q)\g H}$ and the projection
operator
\begin{equation}
 \left( \begin{array}{cc} Q &\sqrt{Q(\idty-Q)}\Bigr|_{\g K} \\ 
 \sqrt{Q(\idty-Q)} &(\idty-Q)\Bigr|_{\g K} \end{array}\right)
\label{purification}
\end{equation}
on $\g H\oplus\g K$.
For the homogeneous states $Q=\kappa\idty$ with $0<\kappa<1$, $\g K=\g H$
and the projector becomes
\begin{equation*}
\left( \begin{array}{cc} \kappa\idty &\sqrt{\kappa(1-\kappa)}\idty
 \\ \sqrt{\kappa(1-\kappa)}\idty&(1-\kappa)\idty
\end{array}\right)
\end{equation*}   

For a symbol $Q$ of finite rank, the entropy of $\omega_Q$ is given by
\begin{equation}
 S(\omega_Q) = \tr \Bigl(\eta(Q)+\eta(\idty-Q)\Bigr).
\label{stateentropy}
\end{equation} 
The formula can obviously be extended to compact $Q$ with eigenvalues
converging sufficiently fast to 0. 

In order to compute the dynamical entropy for quasi-free evolutions with a
quasi-free reference state, it suffices to consider a restricted class of
partitions $\c X$ characterised by the property that
\begin{equation*}
 y\mapsto \Lambda(y):=\sum_j x^*_jy\,x_j
\end{equation*} 
transforms the gauge-invariant quasi-free states into themselves. Such maps
$\Lambda$ are called gauge-invariant quasi-free completely positive maps
and are determined by two linear operators $V$ and $W$ on $\g H$
obeying the restrictions
\begin{equation*}
 0\le W\le \idty-V^*V.
\end{equation*}
On a monomial $\Lambda$ acts as
\begin{equation*}
 \Lambda(a^{\#}(f_1)\cdots a^{\#}(f_n)) = \sum_{S\subset\{1,\ldots,n\}}
 \epsilon(S) \Bigl(\prod_{j\in S} a^{\#}(Vf_j)\Bigr)\, \omega_W\Big(
 \prod_{k\notin S} a^{\#}(f_k)\Bigr). 
\end{equation*}
In this formula, $a^{\#}$ denotes either $a$ or $a^*$ and $\epsilon(S)$
equals $\pm1$ according to the parity of the permutation defined by $S$. 
The quasi-free state
$\omega_Q$ transforms under $\Lambda$ into the quasi-free state with symbol
\begin{equation}
 V^*Q\,V+W. 
\label{statetransform}
\end{equation} 
Even if different partitions may yield the same map $\Lambda$,
$\En[\omega_Q,\g X]$ will only depend on $Q$ and $\Lambda$ and we shall
derive in the next proposition its expression directly in terms of $Q$ and
$\Lambda$. 

\begin{proposition}
 Let $\dim(\g H)<\infty$ and let $\g X = (x_1,x_2,\ldots,x_n)$ be an
operational partition in $\c A(\g H)$ such that $y \mapsto \sum_j
x^*_jy\,x_j$ is gauge-invariant quasi-free determined by $(V,W)$. Let the
symbol $Q$ determine the gauge-invariant quasi-free state $\omega_Q$, then
\begin{equation}
 \En[\omega_Q,\g X] = S(\omega_R)
\label{partitionentropy}
\end{equation}  
where $R$ is the symbol on $\g H\oplus (Q(\idty-Q)\g H) =: \g H\oplus\g K$ 
given by
\begin{equation}
 R = \left( \begin{array}{cc} V^*Q\,V+W & V^*\sqrt{Q(\idty-Q)}\Bigr|_{\g K} \\
 \sqrt{Q(\idty-Q)}\, V & (\idty-Q)\Bigr|_{\g K} \end{array}\right)
\label{partitionsymbol}
\end{equation}
\end{proposition}

\begin{proof}
 Let us denote by $(\g H_Q,\pi_Q,\Omega_Q)$ the GNS~triple of $\omega_Q$ and by $(e_1,e_2,\ldots,e_n)$ the canonical orthonormal basis of $\Cx^n$. The pure state on
 $\g B(\g H_Q) \otimes \c M_n$ induced by the vector $\sum_j \pi_Q(x_j)\Omega_Q \otimes e_j \in \g H_Q \otimes \Cx^n$ restricts to generally mixed states on $\c M_n$ and $\g B(\g H_Q)$ that have, up to multiplicities of 0, the same spectrum. A straightforward computation shows that this restriction to $\c M_n$ is the correlation matrix $\rho[\g X]$ and that to $\g B(\g H_Q)$ the density matrix 
\begin{equation*}
 \sum_j |\pi_Q(x_j)\Omega_Q\>\<\pi_Q(x_j)\Omega_Q|.
\end{equation*}  
As $\c A(\g H)$ is isomorphic to the algebra $\c M_{2^{\dim(\g K)}}$ 
we can write that $\g H_Q = \Cx^{2^n} \otimes \g L$ with $\pi_Q(x) = x 
\otimes \idty_{\g L}$. Therefore, there exists a unique unity preserving completely positive map $\Gamma$ on $\g B(\g H_Q)$ determined by the requirement
\begin{equation*}
 \Gamma(y \otimes z) := \Bigl(\sum_j \pi_Q(x_j^*)y\,\pi_Q(x_j)\Bigr) \otimes z, \quad y \in \pi_Q(\c A(\g H)),\ z \in \g B(\g K).
\end{equation*}
Obviously $\En[\omega_Q,\g X] = S(|\Omega_Q\>\<\Omega_Q|\circ\Gamma)$. It
now remains to compute this quantity.

Using the purification~(\ref{purification}) we see that the
GNS~representation space of $\omega_Q$ is the Fock space built on $\g H
\oplus (Q(\idty-Q)\g H)$ and that $\Gamma$ is determined by the operators
$(V\oplus\idty,W\oplus0)$. It suffices now to use
formulas~(\ref{statetransform}) and~(\ref{stateentropy}) to finish the
proof.
\end{proof}

We shall now obtain a lower bound on the dynamical entropy in terms of
currents of particles falling into a trap. This will generally not provide
the optimal lower bound but we expect it to provide the correct growth
exponent, which it certainly does in the case of linear growth. The second
quantised version of a localised trap is provided by a quasi-free 
completely positive map $\Lambda$ with operators $(V,0)$. In order to
compute the number $\Delta$ of particles that disappear from an
homogeneous  state $\omega_\kappa$ in the trap, we consider the particle
number operator $N$ in $\c A(\g H)$. $N := \sum_j a^*(e_j)a(e_j)$ where
$\{e_1, e_2, \ldots\}$ is an orthonormal basis of $\g H$. We assume for the
moment that $\g H$ is finite dimensional but the general case can be
obtained by a suitable limiting procedure. Then
\begin{equation*}
 \Delta = \omega_\kappa(N - \Lambda(N)) = \kappa \tr (\idty - V^*V).
\end{equation*}  
The locality of the trap is expressed by the condition
\begin{equation*}
 \rank(\idty-V^*V) < \infty,\qquad V^*V \le \idty.
\end{equation*}

In our case, all refined and evolved partitions remain quasi-free with
strictly local action. An explicit computation shows that
$\g X_t$ is determined by $(V_t,0)$ with
\begin{equation*}
 V_t = [VU]^t U^{-t}.
\end{equation*}
Using the explicit expression of the entropy of correlation
matrix~(\ref{partitionentropy}) in terms of its
symbol~(\ref{partitionsymbol}), we have 
\begin{equation}
 \En[\omega_\kappa,\g X_t] = \tr \Bigl(\eta(R_t) + \eta(\idty-R_t) \Bigr) 
\label{entropyrefinement} 
\end{equation}
with
\begin{equation}
 R_t = \left( \begin{array}{cc} 
 \kappa V_t^*V_t &\sqrt{\kappa(1-\kappa)}\, V_t^* \\
 \sqrt{\kappa(1-\kappa)}\, V_t &1-\kappa
 \end{array} \right).
\label{symbolrefinement} 
\end{equation}
In order to avoid a trivial situation we assume that $0<\kappa<1$, in which
case the trace in~(\ref{entropyrefinement}) is taken over a space of
dimension twice the rank of $\idty-V_t^*V_t$. We write
\begin{align*}
 &\left( \begin{array}{cc} 
 \kappa V_t^*V_t &\sqrt{\kappa(1-\kappa)}\, V_t^* \\
 \sqrt{\kappa(1-\kappa)}\, V_t &1-\kappa
 \end{array} \right) \\
 &\qquad= \left( \begin{array}{cc} 
 \sqrt\kappa V_t^* &0 \\
 \sqrt{1-\kappa} &0
 \end{array} \right) 
 \left( \begin{array}{cc} 
 \sqrt\kappa V_t &\sqrt{1-\kappa} \\
 0 &0
 \end{array} \right) =: A^*A.
\end{align*}
But then
\begin{equation*}
 A\,A^* = \left( \begin{array}{cc} 
 1-\kappa +\kappa V_t\,V_t^* &0 \\
 0 &0 \end{array} \right).
\end{equation*}
Using that, up to multiplicities of 0, $A^*A$ and $A\,A^*$ have the same
spectrum and that $\eta(0)=\eta(1)=0$, the entropy becomes
\begin{equation*}
 \En[\omega_\kappa,\g X_t] =
 \tr \eta(1-\kappa(1-V_t^*V_t)) + \eta(\kappa(1-V_t^*V_t)).
\end{equation*}  
Finally, as $\eta$ is concave we obtain the lower bound 

\begin{proposition}
\label{proposition2}
 \begin{equation*}
  \En[\omega_\kappa,\g X_t] \ge \{\eta(\kappa) + \eta(1-\kappa)\}
  \tr (\idty - V_t^*V_t).
 \end{equation*}
\end{proposition}

The usual computation of dynamical entropy involves two more steps. First
the computation of the asymptotic rate of entropy production $\en[\omega,\g
X]$ which consists in taking the limit for $t\to\infty$ of $\En[\omega,\g
X_t]/t$ and next taking the supremum over a suitable class of operational
partitions. Proposition~\ref{proposition2} is more general in the sense
that it provides a lower bound for $\En[\omega,\g X_t]$ even when this
quantity scales in a sublinear way in $t$. We can also use the lower bound
of the proposition to show that the dynamical entropy is strictly positive
whenever there is a non-zero asymptotic current for a trap belonging to the
class of allowed partitions. We have seen that $\kappa \tr (\idty -
V_t^*V_t)$ represents the total amount of particles that disappeared from
the homogeneous state $\omega_\kappa$ into the trap up to time $t$. The
corresponding current at time $t$ is then 
\begin{equation*}
 \kappa \tr (V_t^*V_t-V_{t-1}^*V_{t-1}),
\end{equation*}
which is precisely the quantity considered in Section~\ref{section2}. In
particular, the entropy grows linearly in time if the absolutely continuous
spectral subspace of the single-step unitary $U$ is non-trivial. But even if
$U$ has no absolutely continuous spectral component, an estimate of the
growth exponent of the entropy may be obtained.   
\medskip

\noindent
\textbf{Acknowledgements}
It is a pleasure to thank J.~Quaegebeur and F.~Redig for quite useful
discussions and comments.

\end{document}